# X-ray and Gamma-ray Observations of the Fermi Bubbles and NPS/Loop I Structures

**Jun Kataoka** [1,*], **Yoshiaki Sofue** [2], **Yoshiyuki Inoue** [3], **Masahiro Akita** [1], **Shinya Nakashima** [3] and **Tomonori Totani** [2]

[1] Institute for Science and Engineering, Waseda University, 3-4-1, Okubo, Shinjuku, Tokyo 169-8555, Japan;
[*] Correspondence: kataoka.jun@waseda.jp; Tel.: +81-3-5286-3081
[2] Department of Astronomy, The University of Tokyo, Bunkyo-ku, Tokyo 113-0033, Japan
[3] RIKEN, 2-1 Hirosawa, Wako, Saitama 351-0198, Japan

**Abstract:** The Fermi bubbles were possibly created by large injections of energy into the Galactic Center (GC), either by an active galactic nucleus (AGN) or by nuclear starburst more than ~10 Myr ago. However, the origin of the diffuse gamma-ray emission associated with Loop I, a radio continuum loop spanning across 100° on the sky, is still being debated. The northern-most part of Loop I, known as the North Polar Spur (NPS), is the brightest arm and is even clearly visible in the *ROSAT* X-ray sky map. In this paper, we present a comprehensive review on the X-ray observations of the Fermi bubbles and their possible association with the NPS and Loop I structures. Using uniform analysis of archival *Suzaku* and *Swift* data, we show that X-ray plasma with $kT$ ~ 0.3 keV and low metal abundance ($Z$ ~ 0.2 $Z_\odot$) is ubiquitous in both the bubbles and Loop I and is naturally interpreted as weakly shock-heated Galactic halo gas. However, the observed asymmetry of the X-ray-emitting gas above and below the GC has still not been resolved; it cannot be fully explained by the inclination of the axis of the Fermi bubbles to the Galactic disk normal. We argue that the NPS and Loop I may be asymmetric remnants of a large explosion that occurred before the event that created the Fermi bubbles, and that the soft gamma-ray emission from Loop I may be due to either $\pi^0$ decay of accelerated protons or electron bremsstrahlung.

**Keywords:** Fermi bubbles; North Polar Spur; Loop I; Galactic halo; X-rays

## 1. Introduction

A supermassive black hole, the mass of which ranges from $10^5$ to $10^{10}$ $M_\odot$, can be found in the center of almost all spiral and elliptical galaxies (e.g., [1,2]). However, why only 10% of galaxies have a very bright nucleus, known as an active galactic nucleus (AGN), and why some AGNs have collimated outflows known as jets, sometimes extending over megaparsecs, remain unanswered (e.g., [3]). With few exceptions, only giant elliptical galaxies nesting a supermassive black hole of mass >$10^{8-9}$ $M_\odot$ have powerful jets, suggesting a close connection between the black hole mass and jet production [4,5]. Most spiral galaxies, including our galaxy, do not have direct evidence for powerful AGNs or jets in their centers. Various observations [6] have confirmed the existence of a supermassive black hole (Sgr A*) of mass $4 \times 10^6$ $M_\odot$ at the center of the Milky Way. The total photoluminosity of Sgr A* at the Galactic center (GC) is ~$10^{33-35}$ erg s$^{-1}$, which is $10^{9-11}$ times lower than the Eddington luminosity for a black hole of mass $4 \times 10^6$ $M_\odot$. Short flaring activities of luminosity up to a few times $10^{39}$ erg s$^{-1}$ have been reported by X-ray satellite missions [7], but that value is still far from the output of powerful, high-luminosity AGNs like quasars at cosmological distances.

Even though the current activity of Sgr A* is low, the evidence of past activity of our GC includes a variety of features. X-ray observations have found the Fe$K_\alpha$ echo from molecular clouds situated a few hundred parsecs apart from and around Sgr A* (see [8] for a review). Moreover, X-ray data from the *Suzaku* satellite found a diffuse overionized clump with a jet-like structure south of the GC, i.e., ~200 pc from Sgr A*, suggesting that an energetic ejection of plasma from Sgr A* occurred about 1 Myr ago [9]. Both X-ray observations can be understood if Sgr A* was much brighter in the past. In particular, in the latter case, the luminosity required to create the overionized clump via photoionization would have to be ~$10^{44}$ erg s$^{-1}$, which is close to the Eddington luminosity. In addition, in the radio sky, Loop I, a continuum loop spanning across 100° on the sky, and its brightest arm, known as the North Polar Spur (NPS), are clearly visible. In the past, there had been hopes of finding similar radio arcs denoted as Loops II, III, and IV [10]. Most researchers believed that these giant structures are created by local old supernova remnants. Amongst them Loop I was thought to be a superbubble in the Sco-Cen star-forming region 130 pc from the Sun [10,11]. In the 1970s, an alternative interpretation was proposed that Loop I and NPS are remnants of a starburst or a nuclear outburst that happened in the GC about 15 Myr ago [12-16]. However, this model has been almost ignored until the launch of the *Fermi* gamma-ray space telescope [17-19].

The Fermi bubbles are gigantic gamma-ray structures recently found by the Large Area Telescope (LAT) [20] onboard the *Fermi* satellite. Some argue that the bubbles were created by a large injection of energy, perhaps from an AGN-like outburst (e.g., [21-24]) or from nuclear starburst (e.g., [25-28]), in the GC. The bubbles extend about 50° (or 8.5 kpc) above and below the GC, have a longitudinal width of ~40°, and are almost symmetrical. The gamma-ray emission of the bubbles spatially correlates with the so-called *WMAP* (Wilkinson Microwave Anisotropy Probe) microwave haze [29], which is spherical with a radius of ~4 kpc centered at the GC; this correlation was recently confirmed by *Planck* observations [30]. Moreover, the recently discovered linearly polarized giant radio lobes emanating from the GC closely correspond to the Fermi bubbles [31]. With these new findings, the NPS/Loop I structure has reentered the spotlight, particularly because the estimated energy (~$10^{55-56}$ erg) and timescale (~10 Myr) needed to create such structures are consistent with those required to create the Fermi bubbles. In this context, the bright X-ray enhancement associated with the NPS, clearly seen in the *ROSAT* all-sky map [32], may be Galactic halo gas swept up and weakly heated via shock expansion [33].

With working hypothesis that the NPS/Loop I and the Fermi bubbles are the phenomena having a common origin of some energetic episodes at the GC, careful X-ray investigations of the Galactic halo gas, the NPS, and Loop I are key and their relationship to the Fermi bubbles have been obtained extensively. In particular, archival *Suzaku* and *Swift* data obtained over 100 pointings inside the Fermi bubbles were systematically analyzed to compare the X-ray emissions inside the Fermi bubbles with the characteristics of surrounding Galactic halo gas [34,35]. An absorbed thermal X-ray plasma with $kT$ ~ 0.3 keV and $Z$ ~ 0.2 $Z_\odot$ was found within Galactic longitude $|l| < 20°$ and latitude $5° < |b| < 60°$, covering the entire extent of the Fermi bubbles. However, questions remain, why are the north and south X-ray skies highly asymmetric, even though the Fermi bubbles are symmetric above and below the GC? What makes the difference of the gamma-ray emissions associated with Fermi bubbles and Loop I, and why only Loop I is exceptionally bright in the radio and X-ray skies? In this paper, we try to answer all these questions by summarizing the current knowledge contained in the archival X-ray data on the Fermi bubbles and the NPS/Loop I structure.

## 2. Gamma-ray view of Fermi Bubbles and Loop I

We start with a quick review of the gamma-ray sky observations obtained with the *Fermi* LAT. Figure 1(a) shows the gamma-ray all-sky map reconstructed for $E > 100$ MeV and based on 8 years of accumulated *Fermi* LAT data. In contrast, Figure 1(b) shows the hardness map, reported for the first time in this paper, where the all-sky map reconstructed for $E > 2$ GeV was simply divided by that for $E < 2$ GeV. Except for enhanced individual point sources, the symmetric bubbles above and below the GC are the only structures clearly seen, suggesting that Fermi bubbles have a harder spectrum than other diffuse foreground emissions. Even though this hardness map does not provide any quantitative information on the morphology and spectra of the Fermi bubbles, it clearly highlights the nature of the bubbles without any ambiguity in any chosen diffuse foreground model. Moreover, we can see a significant enhancement of gamma-ray emission in the southeastern part of the bubbles, which was one of the major discoveries reported in Ackemann et al. [19]. In contrast, there are no signs of diffuse structure associated with Loop I, at least in this hardness map. Therefore, if the gamma-ray spectrum of Loop I exists, it may be much softer than that of the Fermi bubbles.

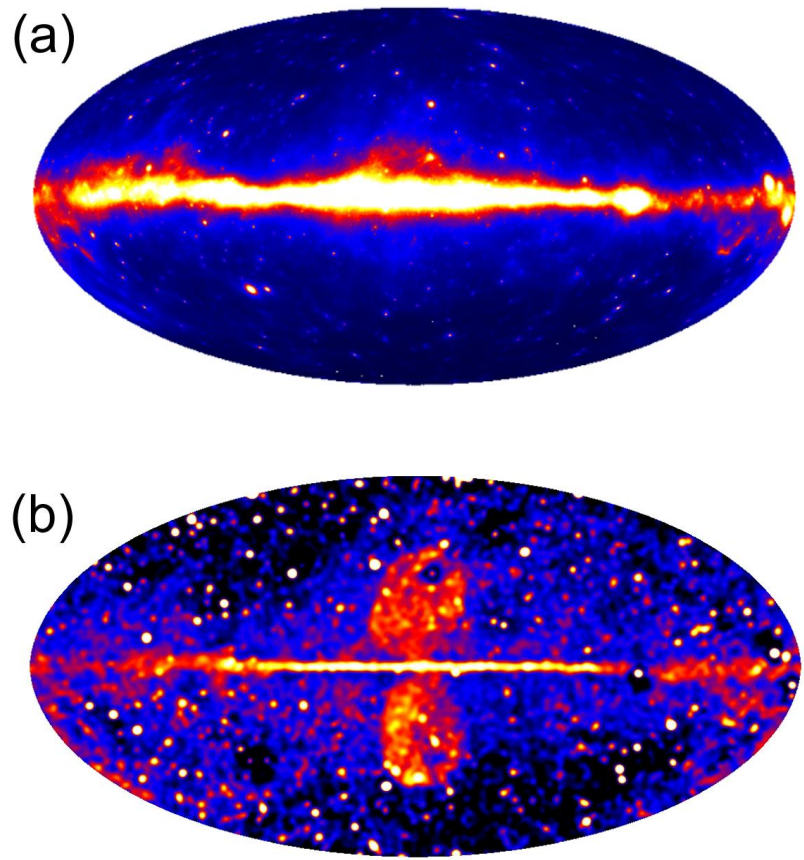

**Figure 1.** (a) *Fermi* LAT all-sky map reconstructed using data from observations above 100 MeV accumulated over 8 years. (b) "Hardness map", in which the high-energy ($E > 2$ GeV) all-sky map was simply divided by the low-energy ($E < 2$ GeV) all-sky map, for the data accumulated over 8 years. The shape of the Fermi bubbles is clearly visible, suggesting that their spectrum is extremely hard compared to that of other foreground emissions.

However, as detailed by Ackermann et al. [19], reliable images and spectra of Fermi bubbles are obtained only after very careful analysis and reduction of *Fermi* LAT data. In summary, the hadronic, inverse Compton, bremsstrahlung, and isotropic extragalactic background radiation comprise the most important foreground emission components to consider in the analysis of large-scale diffuse gamma-ray structures such as Fermi bubbles. These emissions are carefully modeled using the GALPROP cosmic-ray (CR) propagation and interaction code (e.g., [36]), after which the residual emission maps are used to model the Fermi bubbles. Then, the energy spectra of the components are found by simultaneously fitting all the spatial templates to the data. In this process, it is important to estimate the systematic uncertainty in the spectrum of the Fermi bubbles, which is due to the uncertainty in the modeling of the diffuse foreground emissions and the bubbles. From the results, Ackermann et al. concluded that the Fermi bubbles indeed have a very hard spectrum. The power law with an exponential cutoff has a power-law index $\Gamma = 1.9 \pm 0.2$, with cutoff energy $E_{cut} = 110 \pm 50$ GeV. Figure 2 shows the soft and hard residual emission maps for between 700 MeV and 10 GeV [19]. As expected from the hardness map, hard residual emission is mostly dominated by the Fermi bubbles. In contrast, soft residual emission is dominated by the diffuse emission aligned with Loop

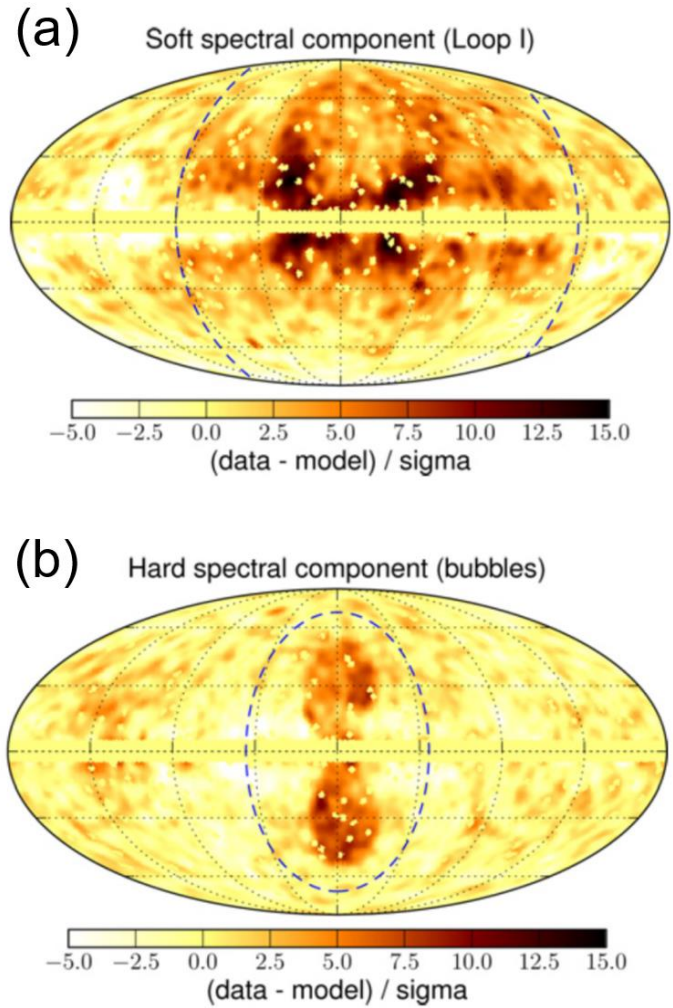

**Figure 2.** Soft and hard spectral residual emission components after subtraction of the hadronic, inverse Compton, bremsstrahlung, and isotropic extragalactic background radiation between 700 MeV and 10 GeV. (a) Soft component is most enhanced around Loop I and its spectrum follows ~$E^{-2.4}$. (b) Hard component shows the shape of the Fermi bubbles and its spectrum follows ~$E^{-1.9}$. (Figure is reconstructed from Figure 13 of [19].

I, for which $\Gamma \approx 2.4$ [19]. This is why Loop I is not seen in the hardness map. The origin of the gamma-ray emission of the Fermi bubbles and Loop I, which may account for the different shapes of the gamma-ray spectra, is revisited and discussed in Section 5.3.

## 3. X-ray view of the NPS

### 3.1. Comparison with the radio morphology

The NPS is a giant structure clearly seen in both the X-ray map and the radio map, with similar morphology to each other. A more detailed comparison of the maps, however, indicates that the radio shell associated with the NPS is sharper than the X-ray shell. In addition, the radio shell is located slightly outside of the bright X-ray shell, as shown in Figure 3. The radio spectrum has a nonthermal power law with a steep differential spectral index $\Gamma \sim 2.5$-3.0, due to the synchrotron emission in origin [37,38]. In contrast, the X-ray spectrum of the NPS is well represented by thermal plasma, as presented in [39,40] and detailed below. Therefore, it is natural to assume that the radio emission of the NPS is the result of a shock front in which radio-emitting electrons are accelerated, whereas the thermal X-ray emission comes from the heated materials swept up by the shock wave just behind it.

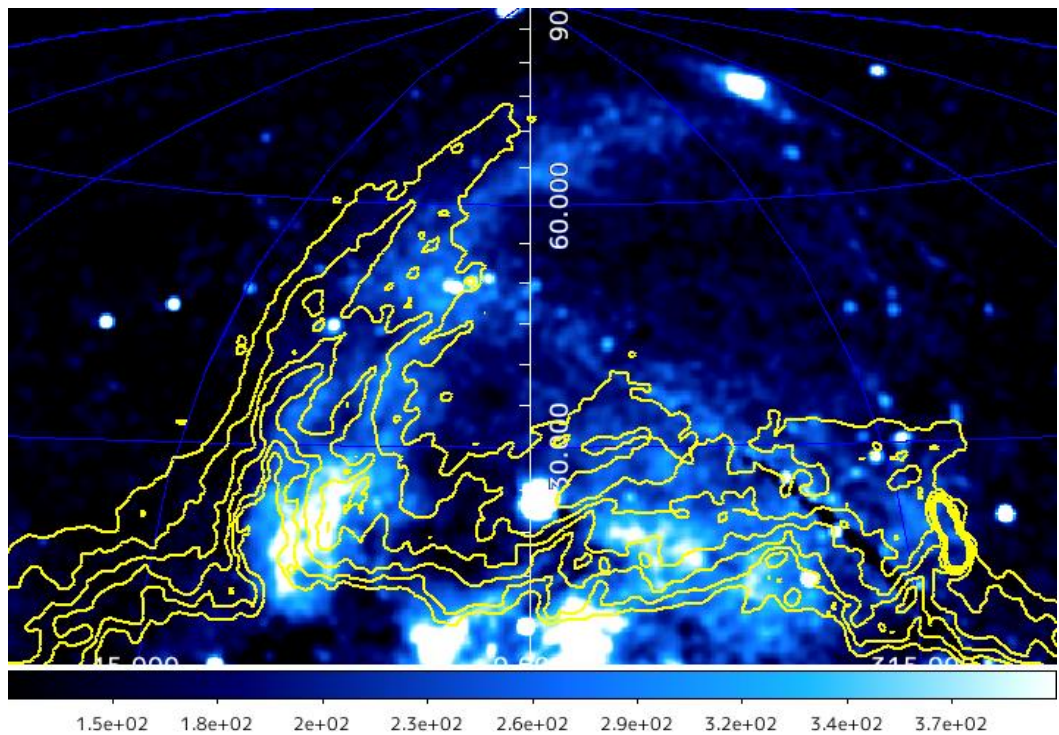

**Figure 3.** (*Blue and white color map*) Close-up of *ROSAT* all-sky map (0.75 keV) around the NPS. (*Yellow contour*) Haslam 408-MHz image of the same region. Note the clear offset between the radio and X-ray maps, where the radio trace of the NPS is slightly outside the X-ray trace of the NPS.

A similar structure consisting of a nonthermal shock front followed by shocked material is observed not only in supernova remnants (SNRs) (e.g., [41,42]) but also in the radio lobe of nearby AGN such as Centaurus A [43]. Therefore, we cannot conclude from its morphology whether the NPS is a close SNR or a shock wave associated with the GC.

Thus, a strong argument against the above GC scenario is based on the measurement of the interstellar polarization at about 100 pc from the Sun, which seems to trace a part of Loop I, including the NPS [44,45]. However, the observed orientation of the stellar polarization is almost perpendicular to the direction of the NPS, especially at low Galactic latitudes, which is at odds with the NPS being associated with a nearby SNR (e.g., [46,47]). Specifically, the NPS radio ridge at $b = 20°–30°$ runs at an angle of 130° (from the GC toward $l = 90°$; [48]), while the optical polarization is at 40°–60° [45]. Thus, this implies that the direction of local magnetic field is nearly perpendicular to the NPS, a scenario that does not support a local SNR as the origin of the NPS. Moreover, if the origin of Loop I (and even Loops II, III, and IV; see [9]) is local, the density of such a giant SNR, with a diameter of ~100 pc, is exceptionally high near the Sun. However, optical filaments, which are often observed in a SNR, have never been seen (e.g., see [49] for the case of the Cygnus loop).

*3.2. Distance to the NPS: 3D view*

Knowing the distance to the NPS is important to fully understand the past activity of our GC and the relationship of the GC to the Fermi bubbles. Recently, analysis of the *ROSAT* archival data showed that the soft X-ray intensity at 0.89 keV along the NPS follows the extinction law due to the interstellar gas in the Aquila Rift, which proves that the NPS is located behind the rift [50]. The mean local standard of rest (LSR) velocity of the Aquila–Serpens molecular clouds is $v = 7.33 ± 1.94$ km s$^{-1}$, which corresponds to a kinematic distance of $r = 0.642 ± 0.174$ kpc. Assuming a shell structure, the lower limit of the distance to the NPS is $1.01 ± 0.25$ kpc, with the center of the shell farther than 1.1 kpc. Moreover, the Faraday distance to the NPS, obtained using the estimated rotation measure (RM), suggests a line-of-sight depth of $r = |RM|/0.82 n_e B$ ~ 5 kpc, where $n_e$ is the electron number density and *B* is the magnetic field strength. Using these measurements, Sofue [50] argued that the NPS is a Galactic halo object, as illustrated in Figure 4.

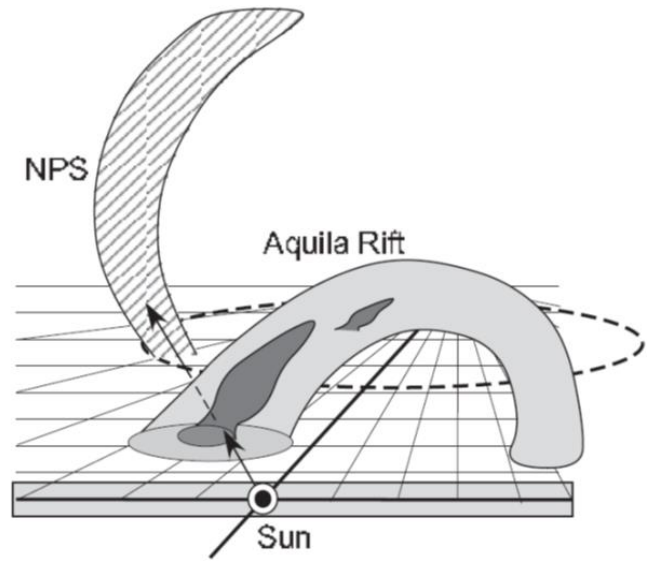

**Figure 4.** Proposed schematic 3D view of the NPS and Aquila Rift, a giant dark lane with a large extinction coefficient that blocks background starlight of the Milky Way, as seen from the Sun. (Figure is reconstructed from Figure 10 of [50].)

Similarly, independent constraints were placed on the distance to the NPS by comparing the foreground interstellar gas column (inferred from X-ray absorption) to the distribution of gas and dust along the line of sight [51]. The columns of X-ray absorbing matter $N_{Habs}$ were derived by spectral fitting the dedicated *XMM-Newton* observations made toward the NPS southern terminus ($l^{II}$ ~29°, $b^{II}$ ~ +5° to +11°). The comparison with X-ray absorption data and local and large-scale dust maps rules out an NPS source near-side closer than 300 pc. The shortest distance to the NPS derived by Lallement et al. [51] clearly demonstrates the absence of a link between the NPS and the nearby Sco-Cen star-forming region, but it supports a possible link between the NPS and the outflow from the GC. Independently, the comparison with the larger-scale Pan-STARRS (PS) 3D dust maps [52] also implies a minimal distance to the NPS of at least 300 pc, which agrees with the evidence in recent studies based on other X-ray data and 3D tomography of dust [53,54].

In contrast, the *Planck* collaboration [30] disfavored a link between the NPS and Fermi bubbles based on the identification of northern and southern polarized emission structures with Loop I secondary arcs and the following points: the strong north-south asymmetry of the NPS, the absence of a pinched structure that is symmetrical above and below the Galactic plane, and the absence of any trace of interaction between NPS/Loop I and the Fermi bubbles. However, NPS/Loop I and the Fermi bubbles may trace completely distinct episodes at different epochs of nuclear activity, in which case the three points become weaker. We revisit this discussion in Section 5.3.

*3.3. X-ray spectra*

The brightest parts of the NPS were targeted by *Suzaku* [39] and *XMM-Newton* [40] for detailed spectral studies. The pointing center of the *Suzaku* observation was (*l*, *b*) = (26.84°, 21.96°), whereas those for three *XMM-Newton* observations were (*l*, *b*) = (25.0°, 20.0°), (20.0°, 30.0°), and (20.0°, 40.0°), respectively, as shown in Figure 5 (*left*). In all observations, the observed X-ray spectra were well represented by the three-component plasma model: APEC1 + Wabs × (APEC2 + PL), where Wabs represents the Galactic absorption as a function of neutral hydrogen column density ($N_H$), APEC1 is an unabsorbed thermal component that represents the Local Bubble (LB) emission, the contamination from the solar-wind charge exchange (SWCX; [55]), or both; APEC2 is an absorbed thermal component that represents the NPS, and an absorbed power-law component (PL) that corresponds to the isotropic cosmic X-ray background (CXB) radiation. Because knowledge of the temperature and abundance of the LB plasma is still poor, $kT = 0.1$ keV and $Z = Z_\odot$ are assumed in both cases. In addition, the photon index for the CXB component is fixed at $\Gamma_{CXB} = 1.41$ (e.g., [56]). In all *Suzaku* and *XMM-Newton* observations, $kT$ ~ 0.25-0.29 keV is obtained for APEC2. Moreover, a depleted abundance $Z < 0.5 Z_\odot$ is also suggested, although enhanced N abundance was reported for the *Suzaku* data [39]. An example X-ray spectrum obtained by *Suzaku* is shown in Figure 5 (*right*).

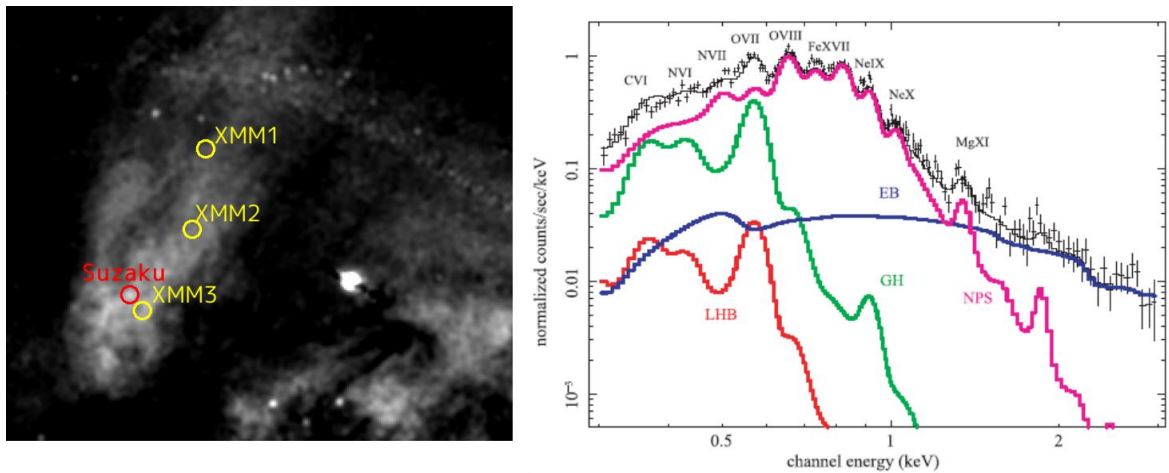

**Figure 5.** (*left*) Pointing centers of a deep (~40 ks) *Suzaku* observation [39] and three short (~15 ks) *XMM-Newton* observations [40] of the NPS. (*right*) Sample X-ray spectrum obtained with *Suzaku* [39]. In this paper, local hot bubble (LHB: *red*) and Galactic halo (GH: *green*) emission are modeled as $kT$~0.1-keV thin thermal plasma (APEC) and solar abundance. Cosmic X-ray background (CXB) is shown as EB (*blue*). The X-ray emission from the NPS is well represented by APEC with $kT$ ~ 0.29 keV and subsolar abundance, except for a detected overabundance of N.

Both the *Suzaku* and the *XMM-Newton* observations that targeted any part of the NPS found a relatively large neutral hydrogen column density ($N_H$), thus substantial amount of absorption is required to model the NPS thermal spectrum in all cases. In fact, Miller et al. [39] reported that the NPS had either >0.71 or >0.97 times the Galactic value of $N_H$ ($N_{H,Gal}$) depending on the choice of background regions, whereas 0.9, 0.6, and 0.5 times $N_{H,Gal}$ was suggested by Willingale et al. [40]. Again, such high $N_H$ values (i.e., the column density was more than 0.5 times the *total* Galactic value in the line-of-sight) support the idea that the NPS is a distant structure near the GC. Willingale et al. [40] argued that the halo and NPS components lie behind at least 50% of the line-of-sight cold gas for which the total Galactic column density is in the range of $(2–8) \times 10^{20}$ cm$^{-2}$. They attributed this high $N_H$ to the cold gas distribution in a wall located at 15–60 pc from the Sun, between the LB and the NPS. However, the presence of such a wall was an assumption made by Willingale et al. [40] so as not to conflict with the observation. Similarly, a high $N_H$ value in the *Suzaku* data was reported [39], but there was no discussion on how to account for the origin of such a large amount of cold gas. Miller et al. assumed throughout their article that the NPS is a local structure based on the interstellar polarization feature and the HI features, both of which cannot be used to strongly support the local interpretation, as we previously discussed.

## 4. Galactic Halo, Fermi Bubbles and Loop-I as seen in X-ray

### 4.1. Galactic halo

If the NPS and other prominent structures near the Fermi bubbles are all related by origin, the Galactic halo is important as a reservoir of thermal gas into which bubbles expand. Although nearby spiral galaxies sometimes exhibit a diffuse thermal X-ray halo that has $kT$ ~ 0.1-0.6 keV and extends out ~10 kpc (e.g., [57,58]), the structural properties, temperature, and metallicity of the Galactic halo gas are not well constrained, which leads to differing observations. For example, Yoshino et al. [59]

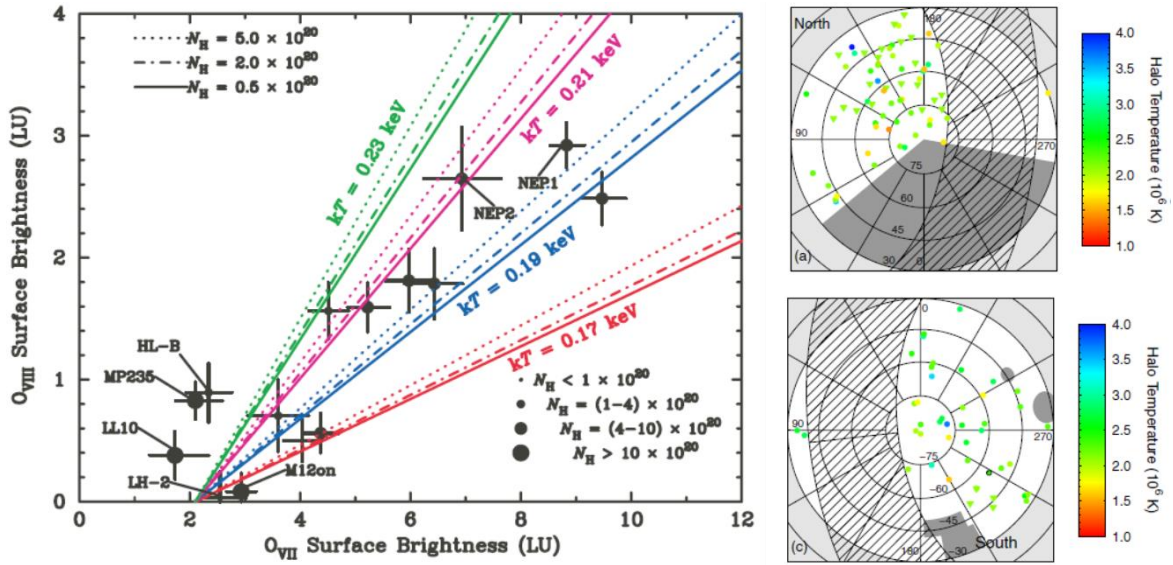

**Figure 6.** (*left*) Relationship between the O$_{VII}$ and O$_{VIII}$ surface brightness of the Galactic halo observed with *Suzaku* for 14 sky fields in 65° < *l* < 295° [59]. The plasma temperature is well constrained in a narrow range of 0.17 keV < *kT* < 0.23 keV. (*right*) Zenith equal-area map showing the temperature of the Galactic halo observed with *XMM-Newton* for 110 sky fields [61]. Again, the Galactic halo temperature is well constrained to *kT* ~ 0.2 keV.

performed a uniform analysis of the diffuse soft X-ray emission associated with the Galactic halo observed in 14 fields by *Suzaku*. They analyzed data obtained within the Galactic longitude range of 65° < *l* < 295° to avoid contributions from the very bright field near the Galactic center (i.e., regions that include the Fermi bubbles were not included). As shown in Figure 6 (*left*), by using the O$_{VII}$ and O$_{VIII}$ line intensities, the authors found that temperatures averaged over different lines of sight are narrowly distributed around *kT* ~ 0.2 keV. Similarly, Henley et al. [60,61] conducted 110 *XMM-Newton* observations of Galactic halo emission and found that the temperature was uniform (median *kT* = 0.19 keV, interquartile range = 0.05 keV), while the emission and intrinsic 0.5–2.0-keV surface brightness varied by over an order of magnitude [(0.4–7) × 10$^{-3}$ cm$^{-6}$ pc] [see Figure 6 (*right*)]. In addition, a "shadowing cloud" was used to test the halo emission, and the same conclusion was reached [62].

In addition, *XMM-Newton* Reflection Grating Spectrometer archival data were used to analyze O$_{VII}$ K$_\alpha$ absorption line strengths in the sightline of 26 AGNs, LMC X-3, and two Galactic sources (4U 1820-30 and X1735-444) [63]. A hydrostatic isothermal model (i.e., King profile or $\beta$ model; [64,65]) was assumed:

$$n(r) = n_0 \left[1 + \left(\frac{r}{r_c}\right)^2\right]^{-3\beta/2}, \qquad (1)$$

where *r* is the distance from the GC, *n*(*r*) is the gas density (in cm$^{-3}$) at *r*, $n_0$ is the density at *r* = 0 (i.e., the Galactic Center), $r_c$ is the core radius, and $\beta$ is the slope of the profile at large radii. The best-fit parameters derived were $n_0 = 0.46^{+0.74}_{-0.35}$ cm$^{-3}$, $r_c = 0.35^{+0.29}_{-0.27}$ kpc, and $\beta = 0.71^{+0.13}_{-0.14}$. These parameters yield halo masses between $M(18\text{ kpc}) = 7.5^{+22.0}_{-4.6} \times 10^8$ M$_\odot$ and $M(200\text{ kpc}) = 3.8^{+6.0}_{-0.5} \times 10^{10}$ M$_\odot$. Although Miller and Bregman did not constrain the temperature of the Galactic halo gas from the absorption line feature, they assumed *kT* ~ 0.11 keV, which they revised in their subsequent papers [66].

Recently, new results from X-ray archival data obtained by *Suzaku* observations of the Galactic halo were reported [67]. The high sensitivity of *Suzaku* to diffuse soft X-ray sources allowed precise determination of the parameters of the hot halo at various lines of sight. The parameters reveal the dependence of the emission measure (EM) of the diffuse plasma on the Galactic latitude, which was not apparent with parameters from previous *XMM-Newton* observations. The obtained temperature of the plasma, *kT* ~ 0.2 keV, was almost constant along the Galactic latitude and longitude, which was consistent with the results of previous works (e.g., [59-62]). There are two possible halo emission density profile models. One is the spherical geometry model, as tested by Miller and Bregman [66]:

$$n(r) = n_0 \left(\frac{r_c}{r}\right)^{3\beta}, \qquad (2)$$

and the other model is the plane-parallel disk-like model [68]:

$$n(z) = n_0 \exp\left(-\frac{z}{h_n \xi}\right), \qquad (3)$$

where *z* is the vertical distance from the Galactic plane, $n_0$ is the density at the Galactic plane, $h_n$ is the scale height, and ξ is the volume filling factor. By fitting the observed EM to the two models, Nakashima et al. [67] argued that the plane-parallel disk-like morphology is preferred over the spherically symmetric morphology for the hot halo.

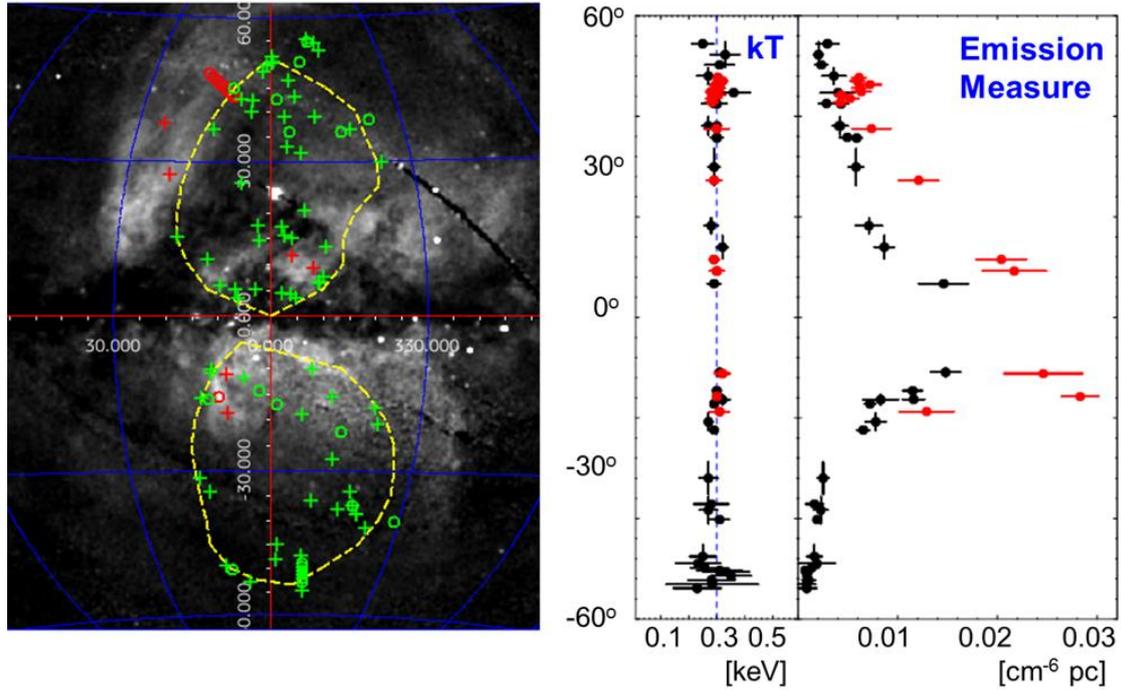

**Figure 7.** (*left*) Positions, in Galactic coordinates, of the 29 *Suzaku* (*circles*) and 68 *Swift* (*crosses*) X-ray data fields overlaid on a *ROSAT* 0.75-keV image [34]. Dashed lines indicate the boundary of the Fermi bubbles, as suggested in [17]. (*right*) Variation in the spectra-fitting parameters EM and *kT* for the APEC model. The parameters for the NPS, enhanced ridge, and clumps are shown in red [34].

*4.2. Interaction between halo gas and Fermi bubbles*

With the above a priori knowledge of the Galactic halo gas, Kataoka et al. [34] made the first attempt at linking the Fermi bubbles and Galactic halo gas. Figure 7 presents their systematic and uniform analysis of archival *Suzaku* (29 pointings; 6 newly presented) and *Swift* (68 pointings; 49 newly presented) data within Galactic longitude |*l*| < 20° and latitude 5° < |*b*| < 60°, covering the full extent of the Fermi bubbles. They found that the plasma temperature is constant at *kT* ~ 0.30 ± 0.07 keV, while the EM varies by an order of magnitude, increasing toward the GC (i.e., low |*b*|) with enhancements at the NPS, the SE claw, and the NW clump. Moreover, the EM distribution of the *kT* ~ 0.30-keV plasma is highly asymmetric between the northern and southern bubbles. They compared the observed EM properties with two simple models: (i) a filled halo without bubbles, the gas density of which follows a hydrostatic isothermal model [see Eq.(1)], and (ii) a bubble-in-halo in which two identical bubbles expand into the halo, forming thick shells of swept-in halo gas. The configuration of the bubble-in-halo model is shown in Figure 8. The observed EM distributions along the Galactic latitude of the two models are compared in Figure 9. Kataoka et al. argued that the EM profile in the north ($b > 0°$) favors model ii, whereas that in the south ($b < 0°$) is rather close to model i. However, a weak excess signature is clearly detected in the southern NPS [denoted South Polar Spur (SPS)]. Such asymmetry, if due to the bubbles, cannot be fully explained by the inclination of the axis of the bubbles with respect to the Galactic disk normal, thus suggesting asymmetric outflow due to different environmental and initial conditions.

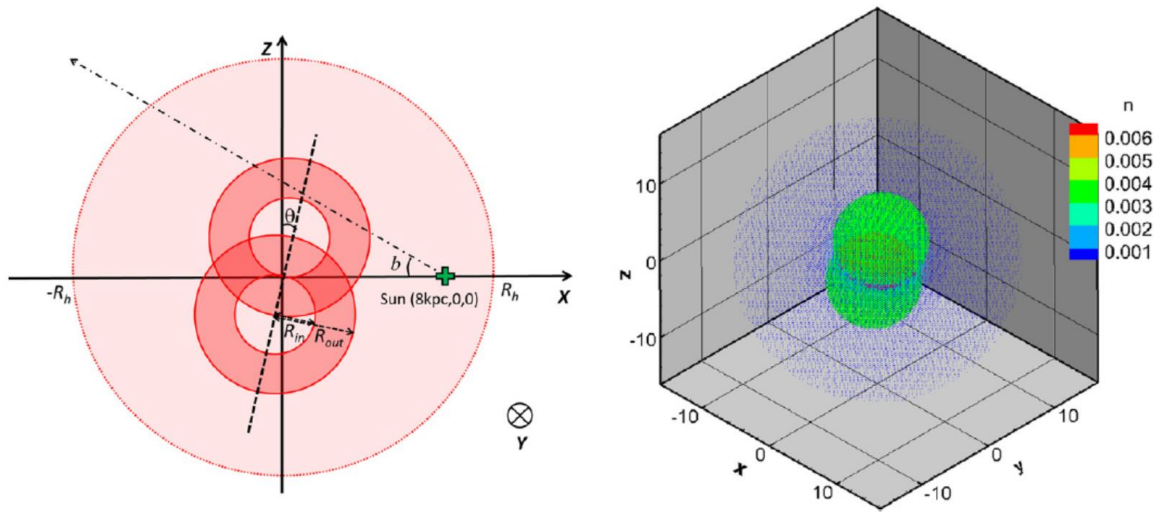

**Figure 8.** Schematic view of bubble-in-halo model proposed in [34]. (*left*) 2D projection at $l = 0°$, where the inner radius is 3 kpc, the outer bubble radius is 5 kpc, and the bubble inclination angle is 10°. (*right*) 3D distribution of gas density profile $n(r)$ in units of cm$^{-3}$. See [34] for more details.

A similar but independent analysis was conducted by Miller and Bregman [69]. They constrained the thermal gas structure of the bubbles by modeling the O$_{VII}$ and O$_{VIII}$ emission line strengths from *XMM-Newton* and *Suzaku* archival data. Their emission model included a hot thermal volume-filled bubble component, cospatial with the gamma-ray region, and a shell of compressed material. They found that a bubble-and-shell model with $n \sim 1 \times 10^{-3}$ cm$^{-3}$ and $kT \sim 0.4$ keV is consistent with the observed O-line intensities. The obtained temperature and expansion rate of the bubbles in [69] were slightly higher than those estimated in [33,34], where $kT \sim 0.3$ keV was proposed. These slight differences are probably due to the different approaches and assumptions adopted by the authors. First, Kataoka et al. [33-35] used overall X-ray spectra to determine the thermal temperature of the plasma, whereas Miller and Bregman [69] used O$_{VII}$ and O$_{VIII}$ intensity ratios to probe the plasma temperature. Second, Kataoka et al. [33,34] assumed an approximate null density of the plasma inside the bubbles, an assumption justified by detailed hydrodynamic simulations (e.g., [70]). In contrast, Miller and Bregman [69] assumed that the bubbles are almost center-filled and that the observed plasma is a mixture of shock-heated material and a volume-filled component associated with the Fermi bubbles. However, both groups of authors agree that the X-ray halo, the original temperature of which was $kT \sim 0.2$ keV, now has a temperature of 0.3-0.4 keV, most probably due to the shock expansion that created the Fermi bubbles.

*4.3. Loop I and Fermi bubbles*

As briefly discussed in Section 1, based on morphology, the NPS is thought to be the brightest arm of Loop I, as seen in the radio and X-ray maps. Therefore, if the NPS is a giant structure in the GC, Loop I is also a remnant of the past activity of the GC, but no detailed X-ray studies have been conducted so far. Figure 10 shows a close-up of the *ROSAT* all-sky map around the NPS and Loop I with the suggested boundary of the northern Fermi bubble [17]. NPS aligns well with the northeast edge of the bubble; however, there is a large "cavity" between Loop I and the northwest edge of the bubble. Therefore, a connection between the Fermi bubbles and Loop I may be a matter of debate and a reason to claim that the NPS and Loop I are unrelated to the Fermi bubbles [30]. In this subsection, we use the analysis of archival *Suzaku* data from within Loop I and the cavity regions and try to determine whether our knowledge of the NPS and the Galactic halo gas is applicable to the Loop I regions.

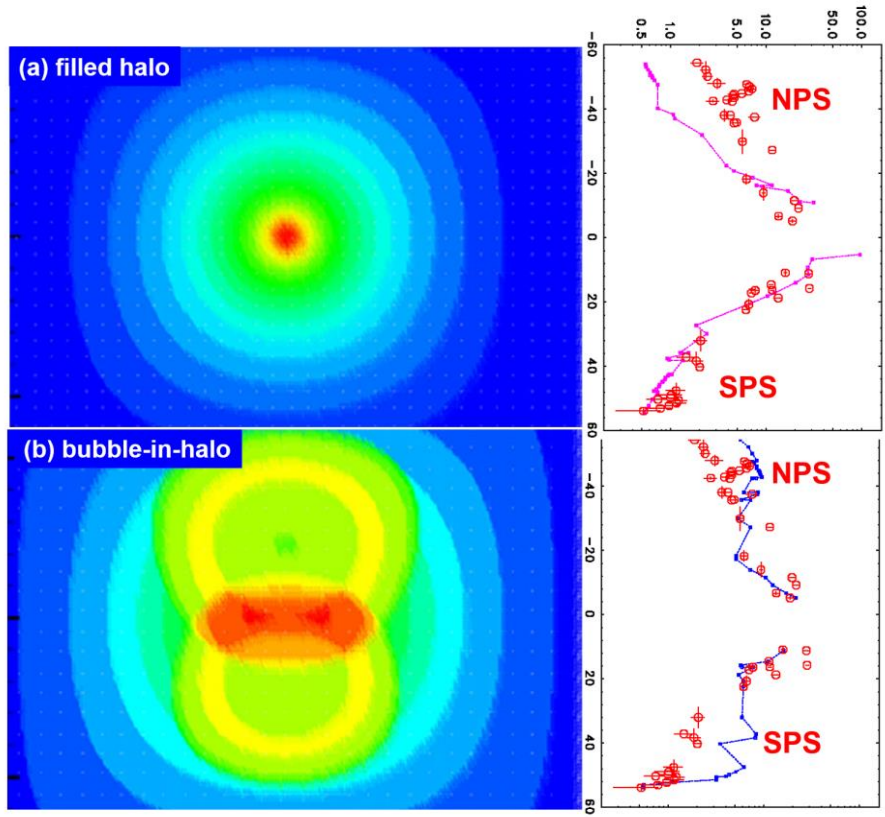

**Figure 9.** (*left*) Variation of EM in the (*l*, *b*) plane as observed from the Sun in the (a) filled-halo model and (b) bubble-in-halo model for inclination angle $\theta$ = 10° [34]. (*right*) Comparison between the observed spectra-fitting parameters as measured using the EM (shown in *red* points) and model predictions (shown as a *magenta* line for the filled-halo model and a *blue* line for the bubble-in-halo model). The NPS is well represented by a bubble-in-halo model but the SPS is not, reflecting the asymmetry between the NPS and the SPS in the X-ray sky maps [34].

We modeled the X-ray spectra of within Loop I and the cavity using the same three-component plasma model as used to reproduce the NPS and Galactic halo, i.e., APEC1 + Wabs × (APEC2 + PL) (see Section 3.3). Figure 11 shows the distribution of EM and $kT$ as a function of Galactic latitude [71]. Clearly, the temperature of the Loop I region is narrowly concentrated around $kT \sim 0.3$ keV, whereas that of the cavity is $kT \sim 0.25$ keV, which is still slightly higher than the average temperature of the Galactic halo. Moreover, obtained $N_H$ is consistent with $N_{H,Gal}$, suggesting that Loop I is also a distant structure in the GC. As detailed in [71], the 30-50% contribution of nonheated Galactic halo gas ($kT \sim 0.2$ keV) compared to the contribution of shock-heated halo gas ($kT \sim 0.3$ keV) along the line of sight explains the temperature of the cavity, whereas both the NPS and Loop I have only a negligible contribution of nonheated gas. In the next section, we consider possible scenarios in which a cavity between Loop I and the edge of the Fermi bubbles is created.

## 5. Discussion

*5.1. Energy and pressure balance between bubble and NPS*

In the spectral fitting of the *Suzaku* data, we did not detect any excess nonthermal emission associated with the bubbles, at least at a level exceeding the expected ~10% fluctuation in the CXB.

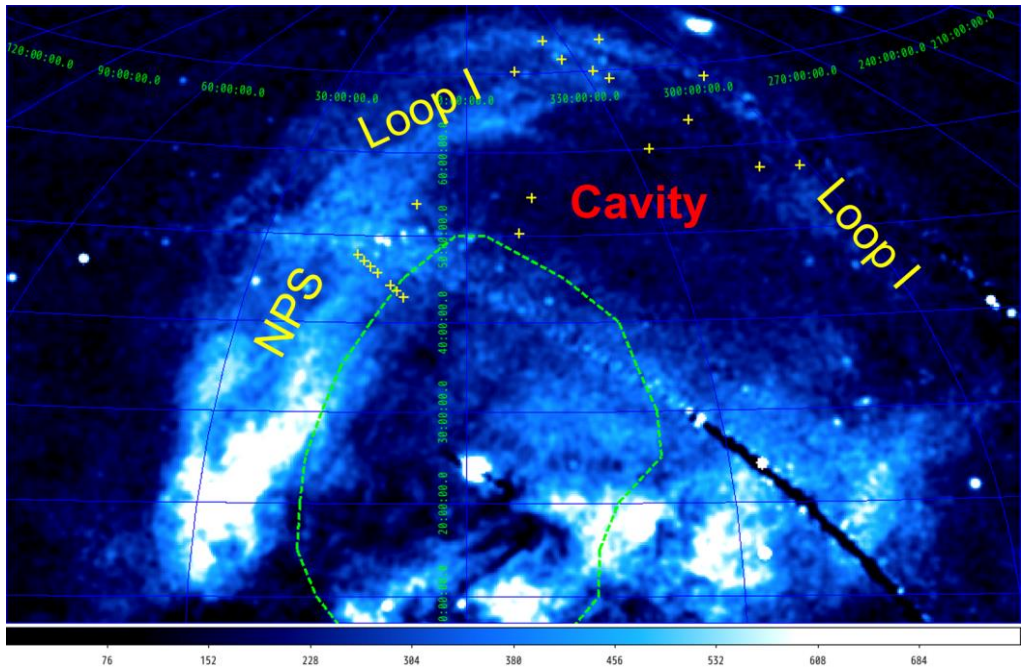

**Figure 10.** *ROSAT* 0.75-keV image showing the positions of the NPS, Loop I, and the northern Fermi bubble as indicated in [17]. While the northeast edge of the bubble aligns well with the NPS, a large cavity exists between Loop I and the northwest edge of the bubble. Yellow crosses indicate the pointing positions of the archival *Suzaku* observations.

Figure 12 shows the spectral energy distribution (SED) of the Fermi bubbles, from radio to GeV gamma ray, with the corresponding upper limit also given in the X-ray band. Since any excess non-thermal emission, at least the level exceeding the fluctuation in the CXB has not been detected for the entire bubbles' region, we derived the upper limit by fitting the spectra assuming an additional power-law component of $\Gamma = 2.0$ for the entire bubbles' region [33] . The GeV data points correspond to the emission of the entire bubbles, following [17]. The radio data points correspond to the *WMAP* haze emission averaged over $b = -20°$ to $-30°$, for $|l| < 10°$; but see also the revised SED in Figure 35 of [19]. A simple one-zone leptonic model in which the radio emission and GeV gamma-ray emission arise from the same population of relativistic electrons through synchrotron and inverse-Compton (IC)) processes of CMB photons is presented in Figure 12 (*blue* curve). If the magnetic field intensity $B = 12$ μG within the bubbles when the emission volume $V = 2[(4/3)\pi R^3]$ for a radius $R = 1.2 \times 10^{22}$ cm, then the nonthermal bubble pressure $p_{n/th} = (U_e + U_B)/3 \sim 2.0 \times 10^{-12}$ dyn cm$^{-2}$, where $U_e$ and $U_B$ are the electron and magnetic field energy densities. The total nonthermal energy stored in electrons and the magnetic field is defined as $E_{n/th} = (U_e + U_B)V \sim 10^{56}$ erg. The results of our model fit suggest an approximate equilibrium, i.e., $U_B \sim U_e$. For comparison, Ackerman et al. [19] derived a slightly smaller value of $B = 8.4$ μG, and there are independent estimates in the literature of $B = 5$–10 μG [17] and $B = 15$ μG [72]. Therefore, it seems reasonable to assume that $B \sim 10$ μG within the bubbles, as long as the leptonic synchrotron-IC/CMB assumption is valid (see an alternative hadronic model proposed in [25]).

In addition, we estimated the thermal pressure of the NPS gas as $p_{th} \sim n_g kT$, where $n_g$ is the gas number density and $kT$ is the gas temperature. We assumed $kT \sim 0.3$ keV and estimated that $n_g = (EM/d)^{1/2}$, where $d$ is the scale length (thickness) of the X-ray plasma with the given EM. If a thermal X-ray envelope or shell has thickness $d \sim 2$ kpc, then $p_{th} \sim 2 \times 10^{-12}$ dyn cm$^{-2}$ and $E_{th} \sim 10^{56}$ erg. Although all these estimates are based on cover-simplified modeling, the pressure and energy of the nonthermal plasma that fills the Fermi bubbles and the thermal plasma immediately surrounding the bubbles are in approximate equilibrium. This clearly supports the proposal that the NPS is composed of Galactic halo gas heated by a shock wave that is driven by the expanding bubbles [21,33]. Indeed, in such a situation, the pressure equilibrium between the shocked downstream fluids is expected. In the framework of the above interpretation, the Mach number of a shock wave that follows from the observed temperature ratio $kT_+/kT_- $ ~ 0.3 keV/0.2 keV is $M$ ~ 1.5, assuming the adiabatic parameter of the Galactic halo gas to be 5/3. From this, it is assumed that the upstream (unperturbed halo gas) pressure $p_- = 0.8 \times 10^{-12}$ dyn cm$^{-2}$, and the shock wave velocity $v_{sh} \sim Mc_{s-} \sim $ 320 km s$^{-1}$, where $c_{s-}$ ~ 200 km s$^{-1}$ is the upstream sound speed. In contrast, Miller and Bregman [69] suggested a higher velocity of $v_{sh} \sim 490^{+230}_{-77}$ km s$^{-1}$, which corresponds to a slightly higher temperature for the heated plasma of $kT_+$ ~ 0.4 keV.

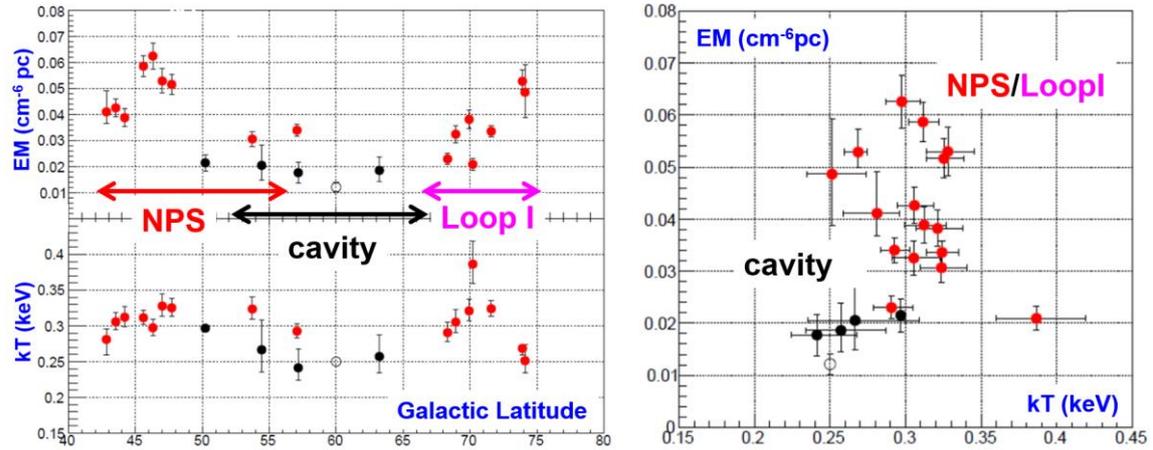

**Figure 11.** (*left*) Distribution of EM and *kT* as a function of Galactic latitude *b* for the NPS, Loop I, and suggested cavity region [71]. (*right*) 2D plot of EM vs. *kT* for the same regions. The NPS and the bright Loop I regions have the same *kT* ~ 0.3 keV, whereas the cavity has a slightly lower *kT* and smaller EM because its emission is dominated by the underlying *kT* ~ 0.2-keV Galactic halo gas.

Next, we present intriguing and independent analyses that used either X-ray or ultraviolet (UV) absorption lines to prove the nonthermal velocity associated with the Fermi bubbles. Fang and Jiang [73] reported an X-ray grating observation of O VII, Ne IV, and O VIII K absorption lines toward the quasar 3C273, which is situated in the Loop I region. They detected a nonthermal projected velocity of 100-150 km s$^{-1}$ and estimated the size of the X-ray absorber to be 5-15 kpc, which is consistent with the volume of the Fermi bubbles. Similarly, Fox et al. [74,75] also reported two high-velocity metal absorption components centered at $v_{LSR}$ = −235 and +250 km s$^{-1}$ in the UV absorption line spectra obtained in the direction of the distant quasar PDS456. These components can be explained if the outflow velocity $v_{out}$ ~ 900 km s$^{-1}$ and the full opening angle is ~ 110°. While $v_{out}$ is higher than what was discussed above, it is dependent on the geometry of the biconical outflow assumed in the model. Moreover, $v_{out}$ does not necessarily coincide with $v_{sh}$. This is evident in most FR II radio galaxies where the jet velocity is relativistic but the expansion velocity of the radio lobes is much slower, typically in the range of ~ 0.001-0.01$c$ (e.g., [76,77]).

*5.2. Comparison with hydrodynamic simulation*

Fermi bubbles are also being actively studied to understand their origin and physical properties. Some bubbles have AGN-like jet activity in the GC (e.g., [21-24]) and others have a starburst outflow (e.g., [26-28]). Although the physical origin of the injected energy is still a matter of debate, metal abundance measurements should provide an important clue to the origin of the Fermi bubbles [76]. In addition, Fermi bubbles are considered a scaled-up version of SNRs and a reservoir of cosmic ray electrons and protons (e.g., [79,80]). In this context, Sofue et al. [70] proposed the bipolar-hypershell (BHS) model for the east and west NPSs (NPS-E, NPS-W, and Loop I) and for the southern spurs SPS-E and SPS-W. The model is based on a numerical hydrodynamic simulation that examines the propagation of shock waves produced by energetic explosive events in the GC. The distribution of soft X-ray brightness on the sky is modeled by thermal emission from high-temperature plasma in the shock-compressed shell while considering the shadowing of the interstellar H I and H II gases. The result of the simulation is compared with the *ROSAT* wide-field X-ray images in R2, 4, and 6 bands.

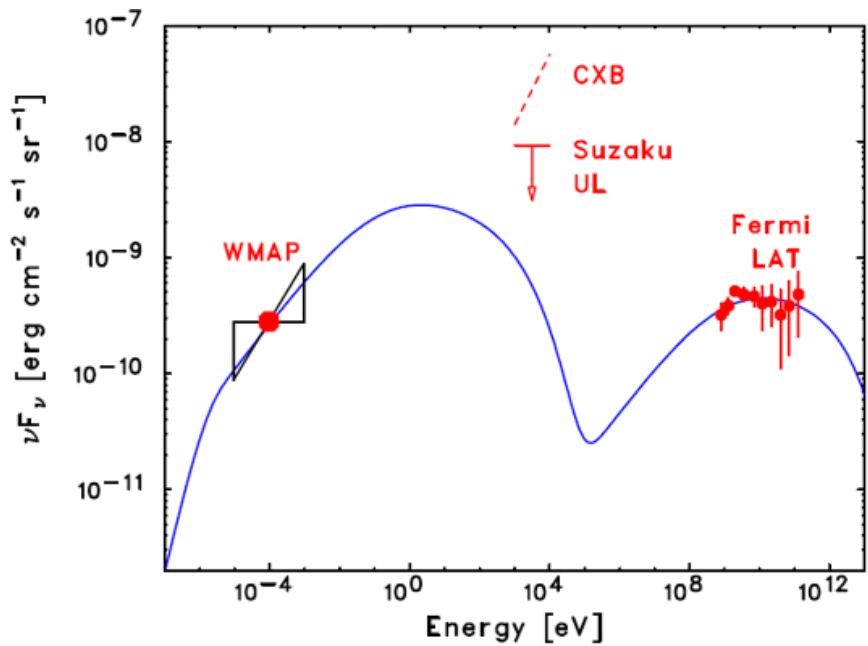

**Figure 12.** SED of the Fermi bubbles fitted with the one-zone synchrotron IC/CMB model (*blue* line). It was assumed that $B = 12$ μG and $R = 1.2 \times 10^{22}$ cm. Given these best-fit parameters, the nonthermal energy stored in the bubbles is $E \sim 10^{56}$ erg at a nonthermal pressure of $p_{n/th} \sim 2 \times 10^{-12}$ dyn cm$^{-2}$, and thus is in approximate equilibrium with the thermal energy and pressure from the NPS. See [33] for more details.

This simulation reproduces the NPS and the southern spurs as shadowed dumbbell-shaped shock waves if a total energy injection of $\sim 4 \times 10^{56}$ erg at the GC occurred $\sim 10$ Myr ago. Figure 13 shows example profiles of gas density, pressure, and temperature measured at $z = 2$ kpc above the GC. For the dense shell that corresponds to the NPS, $kT \sim 0.3$ keV, which is exactly consistent with the observations. Figure 13 shows that the bubbles are filled with plasma with a much higher temperature ($kT \sim 4$ keV), but $n_g$ is one or two orders lower than that of the shock-heated NPS and, thus, does not contribute to the observed X-ray luminosity because for thermal X-ray gas, EM $\propto n_g^2$. Therefore, as long as the BHS model is correct, we can ignore the contribution from the inner bubbles, as is the case in the simple model assumed by Kataoka et al. [34]. Sarkar et al. [81] presented similar but different hydrodynamic simulations of both star formation-driven and black hole accretion-driven wind models. They reached essentially the same conclusion discussed above, where $v_{sh} \sim 300$ km s$^{-1}$ is needed to heat the halo to $kT \sim 0.3$ keV. The corresponding age of the Fermi bubbles is 15–25 Myr. Either a star formation rate of $\sim 0.5$ M$_\odot$ yr$^{-1}$ at the GC or a very-low-luminosity jet and accretion wind arising from the central black hole can produce such an event.

*5.3. Unified idea: Linking NPS, Loop I, and bubbles*

Throughout this paper, we argue that the NPS/Loop I is a giant structure in the Galactic halo and is possibly related to the Fermi bubbles rather than being a local SNR. This idea was proposed first on the basis of radio observations [12], followed by various X-ray observations as discussed above. Here we further comment on the "non-loop" morphology of the Loop I. Careful insights into the all-sky radio and X-ray maps indicate that Loop I is not a complete loop at all, despite of its widely used name after an originally traced small circle on the sky [10]. In fact, only one quarter of the loop in the northeast is traced by the NPS, but the other three quarters are almost invisible, particularly the southern two quarters. Instead, we may trace four gamma-ray spurs emanating from $l \sim \pm 30°$ perpendicularly to the galactic plane, as clearly seen in Figure 2(a), one of which is the NPS at $l \sim +30°$. These four spurs compose two double-horns, symmetric with respect to the galactic plane and the rotation axis of the Galaxy. Also, all the four spurs get brighter toward the galactic plane, which is positionally coincides with the GC, but not with the center of Loop I.

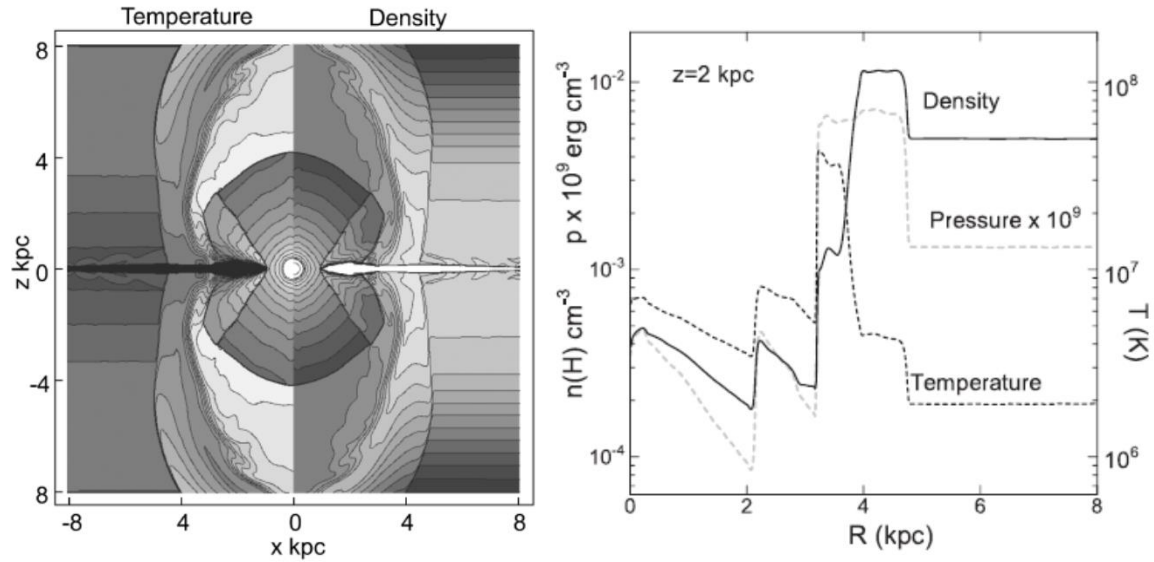

**Figure 13.** (*left*) Density and temperature contours at 10 Myr obtained by hydrodynamic simulation of the GC bipolar-hypershell model as detailed in [70]. Density contours are drawn at log $\rho$ (H cm$^{-3}$) = -4 (*black*) to -1 (*white*) with a dex equal interval of $\Delta \log \rho$ = 0.2, whereas the temperature contours are drawn from log $T$(K) = 5 to 8 with a dex interval of 0.2. (*right*) Density (*solid line*), temperature (*dark gray dashed line*), and pressure (*light gray dashed line*) distributions at constant height $z$ = 2 kpc 10 Myr ago [70].

As for the relationship between the NPS/Loop I and the Fermi bubbles seen in X-ray skies, we showed that (1) the morphology of the NPS seen in the *ROSAT* all-sky map aligns well with the northeast boundary of the bubble. (2) The presence of a large amount of neutral matter, $N_H$, absorbing the X-ray emission of the structure indicates a distance of more than ~1 kpc to the NPS. (3) The $kT$ of the plasma in the NPS is slightly higher than that in other parts of the Galactic halo, suggesting that the NPS is a weakly shock-heated, compressed halo gas. (4) Nonthermal and thermal pressure and energy are in approximate equilibrium between the Fermi bubble and the NPS. However, such close interaction is hardly seen with the southern bubble and halo gas, except for two sharp edges in the south that trace the Fermi bubble below the Galactic disk (e.g., [17,70]), and the only weak sign of the SPS is seen in both radio and X-rays (e.g., [15,33]). This asymmetry between the areas north and south of the GC can be explained by a large-scale outflow from the GC. In fact, most shocked shells, such as SNRs and the GC phenomena, as well as extragalactic jets in the AGN, are more or less asymmetric like the NPS and the SPS. An alternative theory is that the Galactic halo has a structural as well as dynamic asymmetry with respect to the Galactic plane, caused by intergalactic wind (e.g., [14,15]), as discussed in [33].

Therefore, rather than consider the apparent asymmetry of the X-ray halo gas alone, we should explore more seriously why the Fermi bubbles are symmetric while the surrounding spurs are far from symmetrical. Furthermore, the cavity between northwest part of Loop I and the bubble implies an absence of dynamic interaction between the bubble and Loop I, which may contradict the discussion about the NPS and the northeast edge of the bubble. These contradictions are addressed systematically by a two-step explosion process. As an initial condition, we assume that the Galactic halo was extremely asymmetric with respect to the Galactic plane, so much so that the gas density in the northeast area of the halo was enhanced because of the intergalactic wind. The temperature of the Galactic halo was almost uniform and approximated as $kT$ ~ 0.2 keV. The first explosion, either starburst activity or an AGN-like outburst, occurred in the GC about 15-25 Myr ago, releasing $E$ ~ $10^{56-57}$ erg. The expansion velocity of the shock wave was $v_{sh}$ ~ 300 km s$^{-1}$ ($M$ ~ 1.5), which slightly increased the temperature of the Galactic halo gas to $kT$ ~ 0.3 keV, and the gas formed a dense and compressed giant structure like the NPS and Loop I. A corresponding structure (SPS) also formed below the GC, but the SPS is less significant when observed. Then, about 5-10 Myr ago, the second explosion or energetic outflow occurred in the GC and released $E \sim 10^{55-56}$ erg. Since the first explosion had blown away most of the halo gas, the Fermi bubbles that evolved below and above the GC were almost symmetrical. The typical $v_{out} \sim 1000$ km s$^{-1}$, as indicated by the UV absorption line width observed with the Hubble Space Telescope [74,75]. Finally, the NPS and the northeastern bubbles were in contact but left a cavity between the Loop I and the northwest part of the bubble.

Assuming the two-explosion process, gamma rays from the Fermi bubbles are thought to be nonthermal emission from either electrons via the IC/CMB or secondary electrons from accelerated protons, and they would account for the very hard spectrum of $\sim E^{-1.9}$. This corresponds to the electron spectral index $s \sim 2.8$, where the number density of electrons is $N(E_e) \propto E_e^{-s}$. If the same spectrum is modeled with an exponential cutoff power-law, $N(E_e) \propto E_e^{-s} \exp(-E_e/1.3\text{TeV})$, the best-fit index of $s = 2.2$ was obtained. In contrast, gamma-ray emission from the NPS/Loop I structure results from either $p + p \to \pi^0$ decay of accelerated protons or electron bremsstrahlung in the dense, swept-up halo gas at a production rate proportional to $n_g^2$. The observed soft gamma-ray emission of $\sim E^{-2.4}$ suggests that the spectral index of accelerated protons (or electrons) is $s \sim 2.4$. Note that, the spectral indices for electrons and protons may be very close to the well-known spectral index of cosmic rays, supporting the theory that Fermi bubbles are a reservoir of relic cosmic ray electrons and protons (e.g., [25]).

*5.4. Final thoughts*

In this paper, we reviewed the physical origin of the Fermi bubbles, which is still being debated. A leading and fascinating scenario is that past activity in the GC produced outflows such as an AGN jet. As discussed in Section 1, with few exceptions, powerful jets exist only in giant elliptical galaxies, not in spiral galaxies, which suggests a close connection between galactic evolution and jet production. If the nonthermal energy ($E \sim 10^{56}$ erg) stored in the bubble was provided by the jets over 1-10 Myr, the jet luminosity must be $\sim 10^{41-42}$ erg s$^{-1}$. This luminosity is within the minimum luminosity range of low-power extragalactic objects known as FR I radio galaxies [82]. Interestingly, Totani [83] derived similar jet power and energy well before the discovery of the Fermi bubbles. The author argued that various observed features in the GC, including the 511-keV line emission, can be explained within the standard framework of a radiatively inefficient accretion flow in the GC black hole, if the typical accretion rate was about 1000 times higher in the past The outflow energy of such an accretion rate is expected to be $10^{56}$ erg (or $3 \times 10^{41}$ erg s$^{-1}$). Recently, Mou et al. [84] suggested a hydrodynamic model in which the bubbles are inflated by the hot accretion flow.

With respect to morphology, the discovery of gamma-ray emission from the Circinus galaxy is noteworthy [85]. Circinus is a nearby (~4 Mpc) starburst with a heavily obscured Seyfert-type active nucleus, bipolar radio lobes perpendicular to the spiral disk, and kpc-scale jet-like structures. Although the origin of the gamma-ray emission is far from understood, similarities between the Circinus lobes and the Fermi bubbles, including their black hole masses, have been widely discussed. Some indication of similar bubbles perpendicular to the M31 disk has also been reported but is controversial because of the limited photon statistics [86]. These recent observations of similar structures in nearby galaxies imply that AGN-like activity is very common as a specific phase in the evolution of a galaxy, even including quiescent, normal spiral galaxies.

**6. Conclusion**

In this paper, we presented a systematic review of X-ray observations of the Fermi bubbles and the surrounding giant structures NPS and Loop I made as of January 2018. While these structures are generally thought to be a nearby SNR, an alternative interpretation of radio observations in the 1970s claimed that the structures were a remnant of a starburst or a nuclear outburst that occurred near or within the GC about 15 Myr ago. With detailed and uniform X-ray spectral analysis, we added more evidence supporting the theory that the NPS and Loop I are weakly heated Galactic halo gas shocked by a huge explosion. The approximate equilibrium between the nonthermal and thermal gas pressures and energy of the northeast bubble and the NPS is noteworthy. A lingering question, however, is the apparent asymmetry, seen in X-rays, between the NPS and the SPS compared to very symmetric Fermi bubbles. Moreover, a cavity seen in X-ray observations indicates the absence of interaction between Loop I and the northwest part of the bubble. Therefore, we proposed the two-step explosion scenario in which the first explosion occurred in the GC more than 15-25 Myr ago, creating an asymmetric structure like the NPS/Loop I structure that reflected the initial anisotropy of the Galactic halo gas. After the central dense halo gas was swept up, a second explosion occurred in the GC about 5-10 Myr ago to form the symmetric Fermi bubbles. Now the NPS and the bubble edges are in approximate equilibrium, but there is a cavity with no interaction between northwest part of the Loop I and the bubble edge. In this scenario, hard gamma-ray emission from the bubbles is considered to be IC/CMB emission from accelerated electrons, whereas soft gamma-ray emission from Loop I may come from either $\pi^0$ decay of accelerated protons or electron bremsstrahlung and has a spectral index of ~2.4, close to the canonical spectral index of cosmic rays. Future deep observations of the bubbles, the NPS, and Loop I, from radio to gamma rays, will enable further progress toward clarifying the past activity of the GC and its relationship to the Fermi bubbles.

**Acknowledgments:** This review paper was motivated by the exciting workshops that took place soon after the discovery of the Fermi bubbles. The first was held at Stanford University in April 2013, followed by a workshop at Garmisch-Partenkirchen, Germany, in October 2017. In addition, fruitful discussions that took place at the roaming baryon workshop in Sexten, Italy, in July 2017 motivated us to complete this work. We thank Dr. Dmitry Malyshev for useful discussions on the nature of the gamma-ray emission of Loop I and for kindly providing the gamma-ray images presented in this paper that are from [19] or reconstructed from those in [19].